\documentclass[prb,aps,twocolumn,showpacs,superscriptaddress,floatfix]{revtex4}
\usepackage{graphicx}

\begin{document}

\title{Investigations of Optical Coherence Properties in an Erbium-doped Silicate Fiber for Quantum State Storage}
 %\author{Name\corauthref{cor1}\thanksref{label2}}
\author{Matthias U. Staudt}
\affiliation{Group of Applied Physics, University of Geneva,
CH-1211 Geneva 4, Switzerland}
\author{Sara R. Hastings-Simon}
\affiliation{Group of Applied Physics, University of Geneva,
CH-1211 Geneva 4, Switzerland}
\author{Mikael Afzelius}
\affiliation{Group of Applied Physics, University of Geneva,
CH-1211 Geneva 4, Switzerland}
\author{Didier Jaccard}
\affiliation{D\'{e}partment de Physique de la Mati\`{e}re
Condens\'{e}e, University of Geneva, 24 Quai Ernest Anserment,
1211 Geneva 4, Switzerland}
\author{Wolfgang Tittel}
\affiliation{Group of Applied Physics, University of Geneva,
CH-1211 Geneva 4, Switzerland} \affiliation{Institute for Quantum
Information Science, University of Calgary, 2500 University Drive,
NW Calgary, Alberta, Canada, T2N 1N4}
\author{Nicolas Gisin}
\affiliation{Group of Applied Physics, University of Geneva,
CH-1211 Geneva 4, Switzerland} 

\date{\today}

\begin{abstract}
We studied optical coherence properties of the 1.53 $\mu$m
telecommunication transition in an Er$^{3+}$-doped silicate
optical fiber through spectral holeburning and photon echoes. We
find decoherence times of up to 3.8 $\mu$s at a magnetic field of
2.2 Tesla and a temperature of 150 mK. A strong magnetic-field
dependent optical dephasing was observed and is believed to arise
from an interaction between the electronic Er$^{3+}$ spin and the
magnetic moment of tunneling modes in the glass. Furthermore, we
observed fine-structure in the Erbium holeburning spectrum
originating from superhyperfine interaction with $^{27}$Al host
nuclei. Our results show that Er$^{3+}$-doped silicate fibers are
promising material candidates for quantum state storage.
\end{abstract}

%\begin{keyword}
%quantum communication; quantum memory; coherence properties;
%Erbium; rare-earth doped silicate fiber; magnetic tunnelling modes
%PACS: 03.67.HK; 42.50.Md; 32.70.Jz
%\end{keyword}

\maketitle
\section{INTRODUCTION}
%\begin{multicols}{2}
Optical coherence properties of rare-earth-ion-doped
(RE-ion-doped) inorganic crystals have been thoroughly
investigated for photon-echo based optical data storage and data
processing over several decades
\cite{moss82,mitsunaga92a,kroll93}. These materials have
interesting low-temperature properties such as small homogeneous
linewidths (i. e. long coherence times) and large inhomogeneous
broadening (i. e. large frequency bandwidths). Coherent optical
control of RE-ions doped into solids has gained further interest
in recent years as a result of original proposals in the fast
developing fields of quantum communication and quantum computation
\cite{chuang00,pryde00,ohlsson02}. In particular, RE-ions in
crystals are promising for the realization of a reversible
transfer of quantum states between photons and atoms/ions. Such a
quantum memory represents a basic building block for the so-called
quantum repeater \cite{Briegel98}, allowing the extension of
quantum communication schemes such as quantum cryptography
\cite{gisin02} to very long distances.\\An interesting new
approach to a memory for quantum as well as for classical light in
solid state materials is based on controlled reversible
inhomogeneous broadening (CRIB) of a narrow, spectrally isolated
absorption line \cite{kraus05,alex06,nilsson05}. One of the major
challenges in implementing this protocol is the need for an
optically thick atomic medium, with long optical coherence times.
The exact requirements for an efficient CRIB based quantum memory
are still subject to research, but it seems reasonable to assume
that optical depths of a few hundred ($\alpha L >$ 100), and
coherence times longer than a microsecond will be necessary
\cite{extended storage}.\\However, large optical depths and long
coherence times are difficult to obtain simultaneously. Although
coherence times longer than a few ms have been observed
\cite{sellin01}, these demonstrations normally rely on weakly
doped crystals in order to reduce RE ion-ion interactions. Typical
absorption coefficients in weakly doped bulk crystals are of the
order of 1-10 $cm^{-1}$. Because interaction lengths are normally
limited to a few millimeters, by focusing requirements, the
corresponding small optical depths will limit the efficiency of
CRIB-based quantum state storage \cite{alex06}.\\In this article
we investigate the coherence properties of an Er$^{3+}$-doped
silicate fiber at low temperatures, specifically with respect to
the requirements for CRIB. This system is promising for the
realization of CRIB, as arbitrarily large optical depths can be
achieved with low doping concentration through long interaction
lengths. Also, the $^4I_{15/2}\rightarrow ^4I_{13/2}$ transition
in Erbium corresponds to a wavelength of 1532 nm (between the two
lowest crystal-field levels), thus allowing future interfacing of
optical quantum memories with the
standard telecommunication fiber networks.\\
At low temperature, the magnitude and temperature dependence of
the homogeneous linewidth differs significantly between RE-ions in
amorphous disordered materials and in crystalline hosts. The
optical decoherence times are much shorter and quasi-linearly
dependent on the temperature. These properties of amorphous
systems, for example glasses, are due to coupling between the
RE-ion impurity and a broad distribution of low-frequency
tunneling modes (two-level systems or TLS), which are normally not
present in crystals. Although much research has been done to
better understand the low-temperature properties of RE-ion-doped
glasses \cite{huber84,broer86,geva97}, there are still open
questions concerning the exact temperature dependence of the
homogeneous linewidth and its origin. In recent photon-echo
experiments in a Er$^{3+}$-doped silicate fiber
\cite{macfarlane06}, a new magnetic-field-dependent optical
dephasing process was observed. It was proposed that the
field-dependent dephasing is caused by coupling of the large
electronic spin of Erbium to TLS modes of magnetic character.
Magnetic TLS modes have also been considered in the field of
glassy low-temperature physics, including dielectric echo
experiments \cite{vernier93,ludwig02,wuerger02,akbari05}.

\section{EXPERIMENTAL}
 We investigated a single-mode Er$^{3+}$-doped silicate glass fiber (ER
407 from INO, Canada), co-doped as follows: Er 0.07 at \%, Al 2.65
at \%, Ge 3.62 at \%. The length of the fiber was 70 cm,
corresponding to an optical depth of $\alpha L$ = 0.9  at 1532 nm
and 4 K. For measurements between temperatures of 10 K and 2.6 K
the fiber was rolled around a copper cylinder (diameter = 4 cm)
and fixed with adhesive tape to ensure a good thermal contact. The
cylinder was fixed to the cold finger of a pulse tube cooler
(Vericold Technologies). The Er$^{3+}$-doped fiber was fusion
spliced to standard single-mode fibers, which were in good thermal
contact with the cold finger. For measurements at temperatures
lower than 2.6 K an identical fiber was installed in a similar
setup in a home-built $^{3}$He/$^{4}$He dilution refrigerator
where it was cooled to temperatures down to 60 mK. A
superconducting magnet inside this refrigerator enabled us to
apply magnetic fields of up to 2 Tesla.\\To measure the
homogeneous linewidth $\Gamma_h$, or the corresponding decoherence
time (lifetime) $T_2=1/\pi\Gamma_h$, over a wide range of
temperatures and external magnetic fields, we performed both
spectral hole burning and photon echo measurements \cite{kaply87}.
The spectral hole burning technique directly yields the
homogeneous linewidth, however, it is limited by laser frequency
fluctuations during the measurement (in our case typically 0.5-1
MHz). Two-pulse photon echoes are generally more robust against
laser frequency fluctuations, if the fluctuations are small
compared to the frequency bandwidths of the optical pulses. The
two-pulse photon echo technique is therefore the preferred
approach for measuring homogeneous linewidths below 1 MHz
(corresponding to more than a 318 ns decoherence time). However,
decoherence times below approximately 100 ns are difficult to
measure, as the requisite short pulses having sufficiently high
peak intensities are difficult to obtain. In addition to these
technical issues, spectral hole burning and two-pulse photon echo
experiments often yield different decoherence times, as spectral
diffusion \cite{broer86,yano92} leads to an apparent
time-dependent homogeneous linewidth. Spectral diffusion has a
larger impact on the linewidth obtained by spectral hole burning,
as such experiments are performed on a longer time scale than
two-pulse photon echoes. The lifetime obtained by two-pulse photon
echo experiments is often referred to as the intrinsic lifetime.
To study decoherence over a wide range of times, the three-pulse
photon echo technique can be used \cite{yano92}. The experiments
that we describe in this article employed all three of these
techniques.

\subsection{SPECTRAL HOLE BURNING EXPERIMENTS}
To measure the homogeneous linewidth $\Gamma_{hom}$ of a
transition in an ensemble of inhomogeneously broadened absorbers,
a laser having a frequency bandwidth small compared to the
homogeneous linewidth is used to selectively excite ions, thereby
creating a spectral hole in the absorption. Its width can be
measured by scanning the laser frequency across its initial value
and measuring the transmitted light intensity. Note that the
$\emph{measurement pulse}$ is typically much weaker than the
\emph{excitation pulse}.
\\The
output from an external-cavity diode cw laser (Toptica DL 100) was
gated by two standard fiber-optic intensity modulators (Avanex SD
20 and IM10) in order to vary the laser intensity between
excitation and measurement. The first modulator created the two
pulses, whereas the second modulator suppressed the laser light
between two consecutive excitation/measurement sequences. A
Stanford delay generator (DG 535) triggered the sequence with a
frequency of 10 Hz, the duration of the excitation pulse being
typically 200 $\mu$s and the measurement time being typically 1 ms
(short compared to the radiative lifetime of $\sim$10 ms). The
light was then coupled into the Er$^{3+}$-doped fiber inside the
refrigerator, and a fiber-coupled photodiode (NewFocus, mod.
2011v) detected the signal at the output of the refrigerator. Note
that the whole setup relied on telecommunication fiber-technology.

\subsection{PHOTON ECHO EXPERIMENTS}In a two-pulse
photon echo (2PE) experiment, a first pulse, ideally a $\pi/2$
pulse, excites the atoms into a coherent superposition of the
ground and excited state, forming an atomic coherence. With time,
inhomogeneous broadening leads to rapid inhomogeneous dephasing.
Application of a second pulse after a time $t_{12}$, ideally a
$\pi$-pulse, reverses the inhomogeneous dephasing, thereby forming
a photon echo a time $t_{12}$ after the second pulse. The
homogeneous dephasing, which cannot be reversed by the second
pulse, leads to an exponential decay of the 2PE peak intensity as
a function of $t_{12}$. A three-pulse photon echo (3PE) can be
seen as an extension of a 2PE, where the second pulse is divided
into two $\pi/2$-pulses, separated by a variable delay $t_{23}$.
In a 3PE the first pulse creates an atomic coherence, but the
second pulse transfers the coherence into a frequency-dependent
population grating in the ground and excited states. The third
pulse scatters off the grating, forming an echo a time $t_{12}$
after the third pulse. In absence of spectral diffusion
\cite{yano92,broer86}, the grating decays with the radiative
lifetime $T_1$, approximately 10 ms in the case of Er$^{3+}$ doped
silicate fibers.
\\To perform 2PE (or 3PE) experiments, we used a setup similar to the
one employed in the SHB experiments. Two (or three) short pulses
were created by the optical intensity modulators (IMs). The two
IMs were synchronized and acted in series in order to have a good
peak-to-background intensity ratio. The excitation pulses had
durations of t$_{2PE}$=13 ns and t$_{3PE}$=20 ns, with a
repetition frequency of 10 Hz. The pulses were then amplified by
an optical amplifier (Variable Gain EDFA, Trillium Photonics, FAM
C1725-A). In order to avoid excessive heating of the refrigerator
due to amplified spontaneous emission from the amplifier, emitted
between consecutive photon echo excitation sequences, we placed an
additional IM between the optical amplifier and the input of the
refrigerator. The photon echo was detected by a fast detector (IR
DC-1 GHz, mod. 1611v, New Focus) after the refrigerator. The
resulting peak powers were in the range of 100-200 mW at the
refrigerator input. Owing to a good extinction of the background
light ($>$ 33 dB), the temperature could be stabilized at 150 mK
and above.

\section{RESULTS}
\subsection{SPECTRAL HOLE BURNING}
\begin{figure}[t]
 \includegraphics[width=0.5\textwidth]{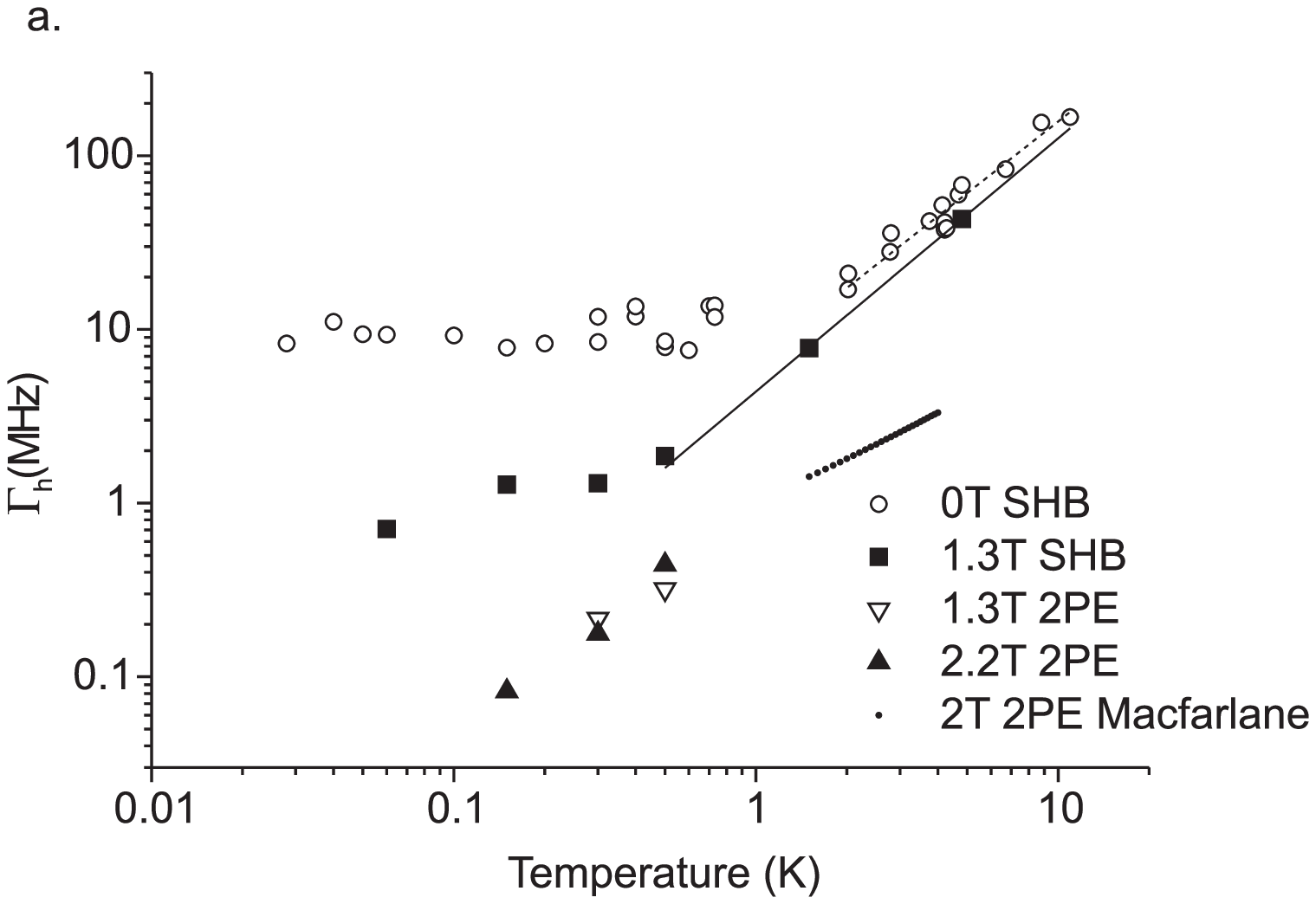}
 \includegraphics[width=0.5\textwidth]{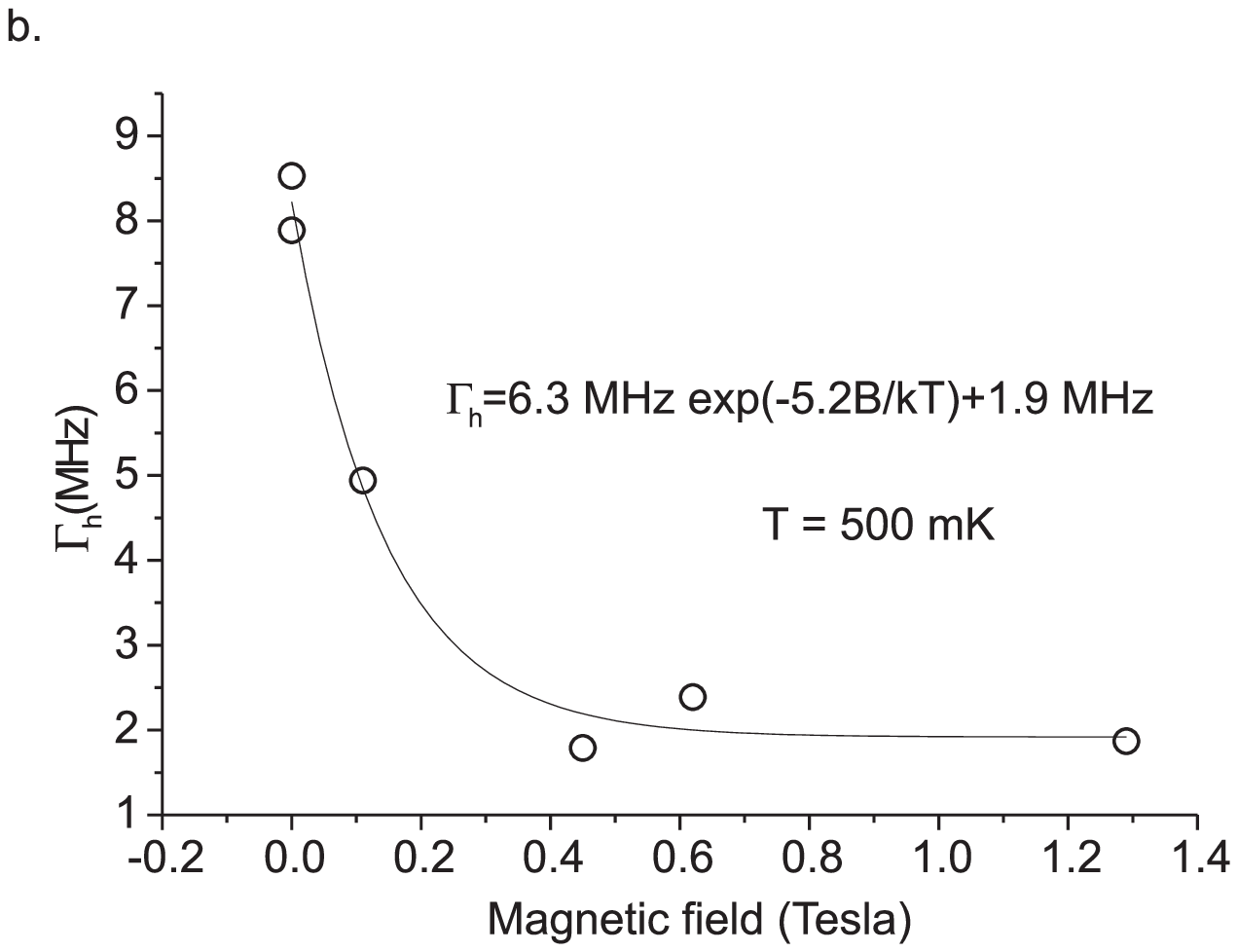}
  \caption{a) Homogeneous linewidth of Er$^{3+}$ as a function of temperature at different
  magnetic fields measured through spectral holeburning and photon echoes. For comparison the data obtained by Macfarlane et al. \cite{macfarlane06} is also plotted. b) Application of
a variable external magnetic field at constant temperature of 500
mK leads to a strong decrease of the linewidth, reaching a
saturated regime at a field of above one Tesla.}
  \label{allT}
  \end{figure}

We measured homogeneous linewidths by SHB at temperatures between
60 mK and 10 K, while applying different magnetic fields ($B$)
ranging from 0 T to 2.2 T (see Fig. \ref{allT}). All linewidths
were extrapolated to zero excitation power, i.e. were corrected
for power broadening effects \cite{mani95}. Based on a large
number of measurements we estimate the error to be of the order of
20 $\%$. Without applied magnetic field, we observed a decreasing
linewidth with decreasing temperature, following a $T^{1.4\pm0.1}$
power-law down to a temperature of 2 K (see Fig. \ref{allT} a
dotted line). Below 2 K the linewidth reached a constant value of
around 8-10 MHz, the slight variations being due to measurement
errors.

\begin{figure}[t]
 \includegraphics[width=0.5\textwidth]{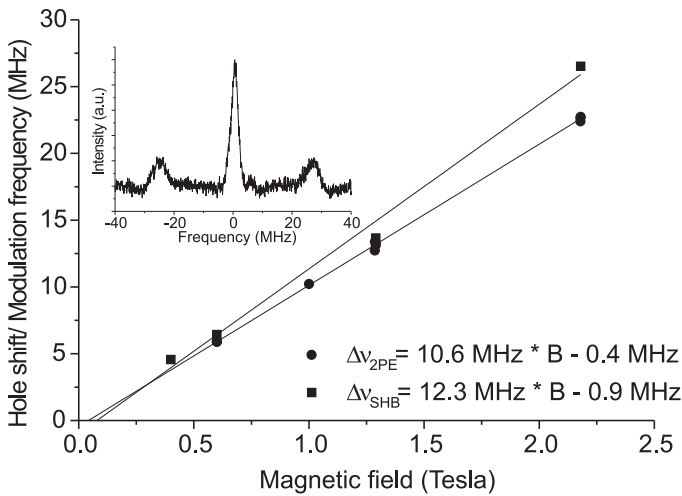}
  \caption{The modulation
frequency of the 2PE decay curve, and the separation between the
central spectral holes and the side holes measured with SHB as a
function of magnetic field. Inset: Spectral hole burning spectrum
in Er$^{3+}$ silicate at a magnetic field of 1.28 T and a
temperature of 60 mK.}\label{afactor}
  \end{figure}

In the case of an applied magnetic field of 1.3 T (see inset Fig.
\ref{afactor}), the transmission spectrum showed not only one
spectral hole at the frequency of the excitation pulse, but also
two side holes equidistant from the central one. We found that the
spectral hole separation was linearly dependent on the magnetic
field (see Fig. \ref{afactor}), with a coefficient of 12.3
MHz/Tesla. We believe that this additional structure in the SHB
spectra are so-called nuclear spin-flip sidebands caused by
coupling between the total angular momentum of the Erbium and
nuclear magnetic moments of nearby $^{27}$Al ions
\cite{wannemacher91}. Due to this superhyperfine, or ligand
hyperfine interaction \cite{abragam70}, optical transitions can
also cause spin flips ($\Delta M=\pm1$) on a neighboring ion.
Therefore one can expect to observe spectral side holes around the
central hole. The observed splitting factor of 12.3$\pm0.8$ MHz/T
can be linked to the $^{27}$Al ion which has a gyromagnetic ratio
of 11.1 MHz/T (for a free, unperturbed ion). The small difference
can be attributed to an enhancement effect \cite{mac84,mac81} in
the silica network. Superhyperfine splitting has also been
observed in other Er$^{3+}$-doped inorganic crystals, eg. YAlO$_3$
(Er-$^{27}$Al), YLiF$_4$ (Er-$^7$Li,$^{19}$F), and LaF$_3$
(Er-$^{19}$F) (see Ref. \cite{wannemacher91}).

The homogeneous linewidth at a magnetic field of 1.3 T, measured
from the spectral hole at the excitation frequency, showed a
temperature dependence similar to that at $B$ = 0 (see Fig.
\ref{allT}a). However, the linewidth continued to decrease
according to a $T^{1.5\pm0.1}$ law down to around 500 mK (solid
line). The measurements at temperatures below 500 mK must be
interpreted with some caution, as laser frequency fluctuations
during the excitation and measurement are likely to play a
significant role. Figure \ref{allT}b shows the homogeneous
linewidths measured at a temperature of 500 mK as a function of
applied magnetic field. We observed a significant decrease of the
linewidth for increasing B field, from 8.2 MHz to a level of
saturation of 2.2 MHz. Note that the analysis of the measurements
at intermediate magnetic fields ($0<B<0.6$ T) was difficult, due
to the three, partly overlapping, spectral holes. We found
similar, strong magnetic-field dependence of the homogeneous
linewidth  at other temperatures below 4 K.

\subsection{PHOTON ECHOES}
\begin{figure}
 \includegraphics[width=0.5\textwidth]{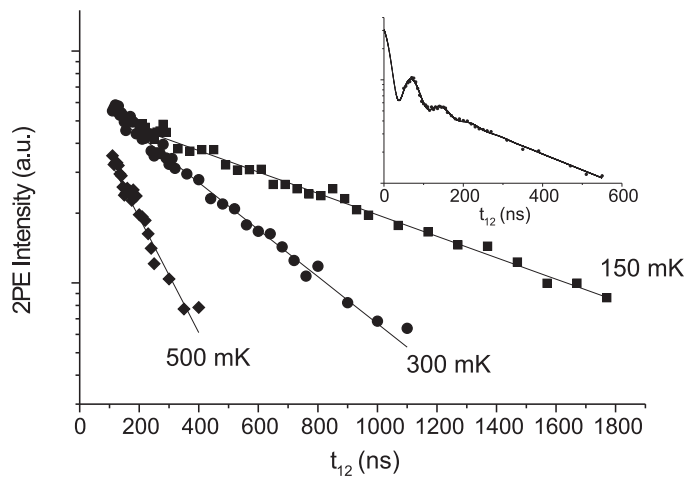}
  \caption{Two-pulse photon
echo peak intensity as a function of time delays between the two
excitation pulses for different temperatures. Inset: For short
time delays a magnetic field dependent modulation was
observed.}\label{echobeat}
  \end{figure}

To measure the intrinsic $T_2$ homogeneous lifetime, we performed
2PE experiments at three sub-Kelvin temperatures, 150 mK, 300 mK
and 500 mK and applied magnetic fields of 0.6 T, 1.3 T and 2.2 T.
In general, one expects to observe a simple exponential decay of
the 2PE peak intensity when varying the time $t_{12}$ between the
two excitation pulses. The associated decay time $\tau$ is
directly connected to the lifetime through the relation $T_2 =
4\tau$. In all our 2PE measurements, however, we observed strong
sinusoidal modulations during the first 100-200 ns of the decay
curve (see inset of Fig. \ref{echobeat}). These modulations can be
explained by coherent excitations of several transitions in the
atoms, as has been observed previously
\cite{wittaker82,mitsunaga92,kim89,kim91}. Although the small
number of modulations visible in our experiments did not allow us
to fit the data to the models proposed in \cite{mitsunaga92}, we
could extract the main modulation frequency. As shown in Fig.
\ref{afactor}, it depends linearly on the magnetic field with a
coefficient of 10.6$\pm0.1$ MHz/T. This value is in close
agreement with the value measured from the spectral hole
splitting, 12.3$\pm0.8$ MHz/T, leading to the conclusion that the
modulation observed with 2PE and the splitting found with SHB have
a common origin. Note that similar modulations have been observed
in Er:Y$_2$SiO$_5$ due to coupling between the Er$^{3+}$ ion and
neighboring $^{89}$Y ions \cite{mac97}. The inset of Fig.
\ref{echobeat} shows that the modulation is strongly damped, which
can be explained by taking into account site-to-site
inhomogeneities in the frequencies of the spin-flip sidebands
\cite{kim89,kim91}.

\begin{figure}
 \includegraphics[width=0.5\textwidth] {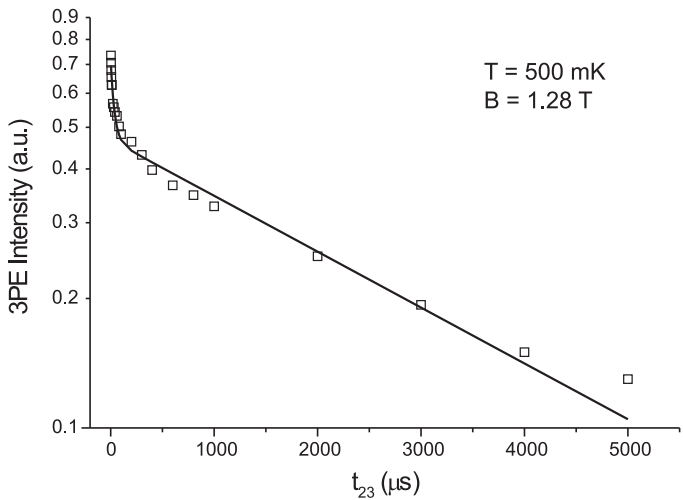}
 \caption{Stimulated
three-pulse photon echo peak intensity as a function of delay time
between the second and the third pulse at a magnetic field of 1.28
T and a temperature of 500 mK.}\label{3pulse}
  \end{figure}

As the modulation of the decay curves disappeared for times
$t_{12}$ longer than 200 ns, we used a simple exponential fit to
compute the decay times based on these data, as shown in Fig.
\ref{echobeat} for a magnetic field of 2.2 T. We observed a strong
dependence of the decay time on the temperature, yielding
lifetimes of $T_2$ = 760$\pm80$ ns, 1710$\pm60$ ns and
3760$\pm140$ ns at temperatures of 500 mK, 300 mK and 150 mK,
respectively. The stated errors refer to two standard deviations,
as obtained from the exponential fit. Measurements with a magnetic
field of 1.3 T and 0.6 T led to similar results. The corresponding
linewidths at 1.3 T and 2.2 T are shown in Fig. \ref{allT}.
Extrapolating these data points to higher temperatures and
comparing them to the SHB measurements at 1.3 T, we find a similar
temperature dependence. Yet, the linewidths obtained through 2PE
are significantly smaller (by a factor of 6 at a temperature of 4
K). This result can be attributed to spectral diffusion that leads
to a time-dependent line broadening. Note that the SHB
measurements below 500 mK are likely to be influenced by laser
frequency fluctuations, as mentioned above, and should therefore
be interpreted with some caution.\\In order to investigate the
role of spectral diffusion, we performed 3PE measurements. The
effect of spectral diffusion on the 3PE process is to reduce the
modulation depth in the population grating by "smearing" it out in
frequency space, thus reducing the efficiency of the echo
formation. In Fig. \ref{3pulse} the 3PE peak intensity is plotted
as a function of the delay time $t_{23}$ between the second and
the third pulse, while the distance between the first two pulses
was kept constant at $t_{12}$=50 ns. A fast decay was observed for
$t_{23}$ shorter than 200 $\mu s$, and a much slower decay for
longer $t_{23}$. As the excited state lifetime is around 10 ms,
much longer than the observed fast decay, this measurement clearly
indicates that spectral diffusion plays an important role on a
microsecond timescale. After 200 $\mu s$, however, the spectral
diffusion reaches a maximum (corresponding to a time-independent
homogeneous linewidth), and the 3PE peak intensity relaxes with a
time constant of around 7 ms, approximately equal to the excited
state lifetime of around 10 ms. A similar behavior has been
observed previously by Broer et al.\cite{broer86} in 3PE
experiments with a Nd$^{3+}$-doped silica fiber.

\section{DISCUSSION}
The temperature dependence of the homogeneous linewidth measured
by SHB at temperatures above 2 K with zero magnetic field, and
above 0.5 K and 1.3 Tesla follows a power-law of $T^{1.4}$ and of
$T^{1.5}$, respectively (see Fig. \ref{allT}a). This can be
explained assuming coupling of the Erbium impurities to TLS modes
\cite{huber84}, that in turn are interacting with the phonon bath.
Flipping of a TLS through phonon interaction leads to a change in
the optical transition frequency of the neighboring RE-ions,
resulting in an optical dephasing. Assuming the standard
approximations in existing models to be correct \cite{geva97} the
temperature dependence for an electric dipole-dipole interaction
between the TLS and the RE-ion should be of the form
$\Gamma_{TLS}^0=aT^{(1+\mu)}$. Here, $\mu$ describes the energy
dependence of the TLS density of states: $\rho(E)\sim E^{\mu}$.
The temperature dependence we found is similar to results
previously obtained through 2PE measurements in an
Er$^{3+}$:silicate fiber ($\Gamma_h\propto T$) \cite{macfarlane06}
and in a Nd$^{3+}$:silica fiber ($\Gamma_h\propto T^{1.3}$)
\cite{broer86}, respectively. The differences can be explained by
a different glass composition, by the use of a different RE-ion,
as has been suggested by \cite{schmidt93}, or by the use of 2PE
measurements as compared to our SHB measurements.\\The
aforementioned model predicts a decrease of linewidth down to
arbitrarily small values when decreasing the temperature. In
contrast to this, our measurements without applied magnetic field
clearly showed the existence of a temperature independent
linewidth below 2K (see Fig. \ref{allT}). Surprisingly,
application of a magnetic field allowed us to reduce the
linewidths by 8-10 MHz. The relative narrowing is particularly
pronounced at temperatures below 2K, but was also observed at a
temperature of 4K. Although a similar, however less pronounced
effect is known to exist in RE crystals \cite{boettger03,mac82},
this behavior has only very recently been observed in a glass by
Macfarlane \textit{et al.} \cite{macfarlane06}. The
field-dependent narrowing cannot be explained by standard models
of coupling of the RE-ion to TLS modes. Furthermore, as has been
pointed out in \cite{macfarlane06} this effect is not likely to
originate from magnetic dipole-dipole interaction between
Er$^{3+}$ ions, as is typical in Er$^{3+}$-doped inorganic
crystals \cite{boettger03,mac82}, since the contribution from
spin-spin interactions is only of the order of 100 kHz at similar
Er$^{3+}$ concentrations. Macfarlane \textit{et al.} instead
proposed that the large electronic spin of Er$^{3+}$ ions couples
to TLS modes having a magnetic character. To describe the
magnetic-field dependent dephasing the authors used a
phenomenological thermal activation law \begin{eqnarray}
\Gamma_h(B,T)=\Gamma_{TLS}^0(T)+\Gamma_{TLS}^1 exp(-g_{eff}\beta
B/kT)\end{eqnarray} where $B$ is the magnetic field, $\beta$ the
Bohr magneton, $g_{eff}$ an effective $g$ factor of the spin
system and $\Gamma_{TLS}^0$ the magnetic field independent TLS
part defined above. Fitting this expression to the homogeneous
linewidths in Fig. \ref{allT}b, we obtained $g_{eff} = 5$ (at 500
mK), similar to the value of $g_{eff}$=3.2 given in
\cite{macfarlane06}. The general features of the magnetic field
dependent linewidth (Fig. \ref{allT}b) correspond well to the
fitted curve. However, due to difficult-to-resolve, partly
overlapping spectral holes at intermediate fields, we estimate the
error in $g_{eff}$ to be around 50$\%$.\\Our results confirm the
recent observation of a strongly field-dependent, optical
dephasing mechanism in RE-ion-doped glasses and can support the
hypothesis of magnetic TLS modes. Note that the magnetic character
of TLS modes has also been proposed to explain unusual magnetic
field dependent dielectric properties of glasses as observed
through dielectric polarization echoes
\cite{ludwig02,wuerger02,akbari05}. Nevertheless, we would like to
point out the possibility that clustering effects in Er$^{3+}$
doped glasses \cite{robin74,ainsle91} might also contribute to the
magnetic-field dependent dephasing. Clustering leads to increased
local Er$^{3+}$ ion concentrations and thereby to enhanced
spin-spin interactions as compared to crystals with similar mean
ion concentration.  Erbium clustering is generally reduced by
codoping with Aluminium \cite{ainsle91}. Yet, the only naturally
abundant isotope of Aluminium  has a strong nuclear spin and is
known to cause strong dephasing in RE-ion-doped inorganic crystals
\cite{yano92}. It would thus be interesting to study the optical
dephasing in fibers having different Erbium concentration, as well
as different concentrations of co-dopants such as Germanium,
Aluminium and Lanthanum.\\The homogeneous lifetimes (or
linewidths) that we measured via 2PE experiments at 0.6 T and
above did not show a magnetic-field dependence, suggesting that a
saturation limit was reached for the temperatures studied (150 mK,
300 mK and 500 mK). This is consistent with the SHB measurements
as a function of magnetic field (see Fib. \ref{allT}b). It should
be noted, that extrapolating our 2PE results to temperatures
between 1.5 and 4K yields homogeneous linewidths of the same order
of magnitude as observed by Macfarlane et al.\cite{macfarlane06}
(see Fig. \ref{allT}a). Yet, in order to compare the coherence
properties of Erbium in the two fibers having slightly different
glass composition, measurements at the
same temperature would be required.\\

An interesting feature of our investigations is the large
difference in homogeneous linewidths measured by 2PE as compared
to the SHB-technique. Due to the resolution limit imposed by laser
frequency fluctuations, this is best exemplified at 500 mK and 1.3
T, where the 2PE and SHB measurements yielded 0.4$\pm 0.04$ and
1.6$\pm 0.3$ MHz, respectively. Note that the second value is
corrected for laser frequency fluctuations. As mentioned earlier
we believe the difference to be due to spectral diffusion, or more
precise spin diffusion, as qualitatively confirmed by the 3PE
measurements shown in Fig. \ref{3pulse}. To quantify the impact of
spin diffusion on the homogeneous linewidth, we analyzed the 3PE
data, taken at 500 mK and 1.3 T, using a spin-diffusion model
previously employed by B\"{o}ttger \textit{et al.}
\cite{boettger03}. The 3PE peak intensity can then be described by
the expression
\begin{eqnarray}\label{3pulseeq}I(t_{12},t_{23})\sim
e^{-2t_{23}/T_1}e^{-4\pi t_{12} \Gamma_h(t_{23})} \end{eqnarray}
where $t_{12}$ is the time difference between the first two
pulses, $t_{23}$ the delay between the second and the third pulse,
and $T_1$ is the excited state relaxation lifetime. In this model
the homogeneous linewidth is expressed as the sum of the intrinsic
part $\Gamma_{0}$ and a part dependent on spectral diffusion,
acting during $t_{23}$ and contributing maximally $\Gamma_{1}/2$ :
\begin{eqnarray}\label{3pulsegamma}\Gamma_h
(t_{23})=\Gamma_{0}+\frac{1}{2}\Gamma_{1}[1-e^{-Rt_{23}}].
\end{eqnarray} Here, $R$ is the rate with which spectral diffusion
takes place. Note that this spectral diffusion model yields a
homogeneous linewidth (in the limit $Rt_{23}\gg 1$) of
$\Gamma_0+\Gamma_1$ if measured through SHB \cite{boettger03}.
Fitting the 3PE peak intensities to this model (see Fig.
\ref{3pulse}) we obtained $\Gamma_1$=1.3$\pm0.1$ MHz,
$R$=0.026$\pm0.005$ $\mu$s$^{-1}$ and $T_1$=6.7$\pm0.5$ ms. If the
2PE experiment is taken to yield the intrinsic homogeneous
linewidth $\Gamma_{0}$ = 0.4 MHz, we can estimate the SHB
linewidth to be $\Gamma_{hom}$=0.4+1.3=1.7$\pm0.1$ MHz. This value
is in surprisingly close agreement with the linewidth of 1.6$\pm
0.3$ MHz that we measured by SHB, supporting the previous
conclusion that spectral diffusion can explain the difference
between the SHB and 2PE results.\\We conclude this discussion by
comparing our results with other measurements carried out in
Er$^{3+}$-doped as well as in other RE-ion-doped glasses. SHB
measurements on various RE-ions in different host materials have
generally resulted in larger homogeneous linewidths as compared to
what is presented in this article (see Fig. \ref{allT}a). For
instance, a linewidth of 106 MHz was measured in a Er$^{3+}$-doped
fluorozirconate glass at 1.6 K \cite{bigot04}. In a
Eu$^{3+}$-doped and a Pr$^{3+}$-doped silicate glass, linewidths
of 24 MHz (1.6 K) and 318 MHz (2 K) were measured \cite{mac83},
respectively. The large differences between all these results,
including ours, can be explained by the presence, or absence, of
an applied magnetic field, different RE-ion concentrations,
different glass compositions, and spectral diffusion.\\Only few
investigations of optical dephasing times through 2PE have been
reported. The 3.8 $\mu$s lifetime that we observed in an
Er$^{3+}$-doped silicate fiber at 150 mK and 2.2 T is, to the best
of our knowledge, the longest lifetime measured so far. Macfarlane
et al.\cite{macfarlane06} recently obtained a lifetime of 230 ns,
at 1.6 K and 3 Tesla, using a similar Er$^{3+}$-doped silicate
fiber. A homogeneous lifetime of 1.6 $\mu$s has been found in a
Nd$^{3+}$-doped silica fiber at 100 mK \cite{broer86}.
Surprisingly, no magnetic field was necessary to obtain this long
lifetime, although Neodymium has an unquenched electronic spin
when doped into a silica glass and therefore strong magnetic
interactions. However, we would like to stress that the fiber was
very weakly doped with Nd$^{3+}$ and had no other co-dopants.
These results indicate that the RE-ion concentration and the
presence and concentration of co-dopants could have an impact on
the magnetic-field dependent dephasing.

\section{Conclusions}We have presented investigations of optical
coherence properties of an Er$^{3+}$-doped silicate fiber. Our
findings reveal a strong magnetic-field dependent optical
dephasing effect that has been discovered only very
recently\cite{macfarlane06}. It is likely that the effect is
linked to tunnelling modes, specific to amorphous systems, which
seem to obtain a magnetic character. For further elucidation more
experimental studies (e.g. spectral holeburning, photon echo and
dielectric echo experiments on fibers having different co-dopants,
RE-ions and concentrations) as well as sound theoretical studies
would be important. Our investigations also revealed, for the
first time in fibers, a superhyperfine structure, which we believe
to be due to
interaction of Er$^{3+}$ ions with neighboring $^{27}$Al nuclei.\\
The fact that $\mu$s coherence times can be achieved at sub-Kelvin
temperatures through application of a magnetic field is promising
for proposals using coherent control of ensemble of ions. For
instance, this would allow for the building of a source of single
photons on demand, based on storage and deterministic recall of a
heralded single photon \cite{fas05}, and a proof-of-principle
demonstration of a quantum memory based on the CRIB protocol.
Furthermore, the coherence time is sufficiently long to transfer
the optical coherences, after absorption of a photonic quantum
state, into superpositions of different ground states with
increased coherence times, thereby increasing the storage time.
The application of a magnetic field also leads to a separation of
the superhyperfine levels as required for the preparation of a
single, homogeneously broadened, absorption line on a
non-absorbing background with sufficiently large frequency
bandwidth. In order to increase the bandwidth beyond the limit set
by the Er$^{3+}$-$^{27}$Al superhyperfine structure, it should
also be possible to use a fiber consisting of a pure
Er$^{3+}$-doped silica core surrounded by depressed index
cladding, provided the
additional levels originate from the presence of $^{27}$Al ions.\\
In conclusion, together with the recent demonstration of
controlled broadening of a transient spectral hole in an identical
fiber \cite{sara06} and the possibility to achieve arbitrary large
optical depths, our results demonstrate the potential of
Er$^{3+}$-doped optical fibers, and RE-doped fibers in general,
for CRIB based quantum state storage.
\section{Acknowledgements}
We would like to thank B. Kraus, M. Nilsson and V. Scarani for
useful discussions. Technical support by C. Barreiro and J.-D.
Gauthier is acknowledged. This work was supported by the Swiss
NCCR Quantum Photonics and by the European Commission under the
Integrated Project Qubit Applications (QAP) funded by the IST
directorate as Contract Number 015848. Additionally M.A.
acknowledges financial support from the Swedish Research Council.

%\end{multicols}


\begin{thebibliography}{99}
\bibitem{extended storage}Note that the memory time can be
extended beyond the optical coherence time by temporarily
transferring the optical coherences to coherences between
different ground state (hyperfine)-levels having longer coherence
times.
\end{thebibliography}
\end{document}